\documentclass[%
 reprint,
 amsmath,amssymb,
 aps,
]{revtex4-2}

\usepackage{graphicx}
\usepackage{dcolumn}
\usepackage{bm}
\usepackage{xcolor}

\begin{document}

\preprint{APS/123-QED}

\title{Enhancing the creation of elements in laser-assisted heavy-ion fusion reactions}

\author{N. Thomson, L. Moschini, A. Diaz-Torres}
\affiliation{%
Department of Physics, Faculty of Engineering and Physical Sciences, University of Surrey, Guildford, Surrey GU2 7XH, United Kingdom
}%




\date{\today}

\begin{abstract}

Low-energy fusion of heavy ions is a fascinating coupling-assisted quantum tunnelling problem, whose understanding is crucial for advancing the synthesis of new elements and isotopes. Quantum dynamical coupled-channels calculations of laser-assisted $^{16}$O + $^{238}$U fusion are presented for both a central collision and the total fusion cross-sections, suggesting that laser-nucleus interaction can enhance the average $^{16}$O + $^{238}$U fusion probability by $6-60 \%$ at subbarrier energies using quasi-static laser fields of intensity $10^{27}-10^{29}$ Wcm$^{-2}$ and photon's energy of $1$ eV. Femtosecond laser pulses are shown to reduce this enhancement by many orders of magnitude.

\end{abstract}

\maketitle


\section{Introduction}

Nuclear fusion is often discussed in the context of green energy production, typically focusing on the interactions of light nuclei such as deuterium ($^2$H) and tritium ($^3$H). Conversely, heavy-ion collisions take energy to fuel the synthesising of super-heavy elements \citep{Armbruster2000,Hofmann2000,Adamian:2020zuo}. The complex structure of interacting heavy ions leads to coupled-channel effects \citep{DASSO,MeasuringBarriers,Hagino2022}, including multi-nucleon transfer reactions which can produce several combinations of elements for the same initial pair of heavy ions \citep{Adamian:2020zuo}.  In this paper, we consider an inert projectile that interacts with a complex heavy-ion target within a fusion reaction. These heavy-ion systems have been discussed extensively in Refs. \citep{DASSO,MeasuringBarriers,Hagino2022,Vockerodt2019,LEE} in varying scenarios. Within a fusion reaction, intrinsic excitations of the heavy ion can be induced. These excitations alter the effective radius of the heavy ion, influencing the probabilities of fusion \citep{Vockerodt2019}. Without a laser field, the effect of these excitations on fusion enhances subbarrier-fusion probabilities, proposing an interesting subject for a laser-nucleus interaction. 

We previously found fusion enhancements in the subbarrier and extreme subbarrier region for a central deuterium-tritium laser-assisted fusion reaction \citep{Thomson}. We therefore extend our model to include couplings of the excited state of the target nucleus to the relative motion of the system in a heavy-ion regime. We again find significant fusion enhancement in the subbarrier region with additional fusion enhancement at energies around the Coulomb barrier. 

This paper is structured as follows: Sect. 2 outlines the quantum dynamical coupled-channels model with subsections dedicated to introducing the laser-nucleus interaction and the fusion cross-sections for the $^{16}$O + $^{238}$U collision as a test case. Sect. 3 and Sect. 4 are the discussion and summary, respectively.

\section{Model} 

We first present the model in the absence of a laser field and subsequently include the laser-nucleus interaction. The heavy-ion collision is described as in Refs. \citep{Vockerodt2019,Vockerodt2}, solving the time-dependent Schr\"odinger equation in the overall center-of-mass reference frame,

\begin{equation}
\begin{split}
i\hbar\frac{\partial\psi_n(r)}{\partial t} = \Big(  -\frac{\hbar^2}{2\mu}\frac{\partial^2}{\partial r^2} +  \frac{\hbar^2 l(l+1)}{2\mu r^2} + V(r) + \epsilon_n 
\\
 +  \, iW(r) \, \Big) \,  \psi_n(r) + \sum_m V_{nm}(r) \, \psi_m(r),
\end{split}
\label{eq:TDSE}
\end{equation}
where the reduced mass  $\mu = m_0 A_t A_p/(A_t + A_p)$ for target and projectile mass numbers $A_t$ and $A_p$ respectively and nucleon mass $m_0$, $l$ is the orbital angular momentum, and $\epsilon_n$ is the $n^{th}$ state excitation energy of the target nucleus. For an $^{16}$O + $^{238}$U interaction, we initially consider the first excited state of the $^{238}$U target, i.e., $n = 0, 1$. It is then extended to an additional excited state ($n = 0,1,2$) including the effects of the $^{238}$U hexadecapole deformation, $\beta_4$.  The indexes $n=0, 1$, and $2$ denote the $0^+, 2^+$ and $4^+$ states of the ground-state rotational band of the $^{238}$U target, respectively. The choice of the $^{238}$U target is due to its low $44.92$ keV first excited state and its large $E2$ reduced transition probability to the ground state \cite{Palffy_2008}. The $^{16}$O projectile will be considered inert due to its high $6.23$ MeV first excited state. 

The elastic channel ($\epsilon_0=0$) Hamiltonian can be extracted from Eq. (\ref{eq:TDSE}) such that,
\begin{equation}
    \hat{\mathcal{H}_0} =  - \frac{\hbar^2}{2\mu}\frac{\partial^2}{\partial r^2} + \frac{\hbar^2 l(l+1)}{2\mu r^2} + V(r) + iW(r).
\label{eq:Hamiltonian}
\end{equation}
The total, two-channels Hamiltonian matrix is then given by,
\begin{equation}
    \hat{\mathcal{H}} = \begin{pmatrix}
    \hat{\mathcal{H}_0} + V_{00}(r) & V_{01}(r) \\
    V_{10}(r) & \hat{\mathcal{H}_0} + V_{11}(r) + \epsilon_1
    \end{pmatrix}.
\label{eq:CouplingHamiltonian}
\end{equation} 
Initially, the wavefunction $\psi_n$ is a Gaussian wavepacket entirely located in the ground state, 
\begin{equation}
    \psi_0 = \mathcal{N}^{-1} \, \exp \left[- \frac{(r-r_{0})^2}{2 \sigma _{0}^{2}}\right] \, e^{-iK_{0}r},
    \label{initialwavepacket}
\end{equation} 
where $\mathcal{N} = \sqrt{\langle \psi_0 | \psi_0 \rangle}$ is the normalisation constant, $r$ is the internuclear radius, $r_0$ is the initial, central position of the wave packet, $\sigma_0$ is the spatial dispersion, and $K_0$ is the average wave number, which depends on the average incident energy $E_0$, $r_0$ and $\sigma_0$ and is found by solving $E_0 = \langle \psi_0 | \hat{\cal H}_0 | \psi_0 \rangle$ using the real part of the Hamiltonian in Eq. (\ref{eq:Hamiltonian}) and neglecting the coupling potential matrix. Eq. (\ref{initialwavepacket}) applies to any partial wave for radial motion because the centrifugal potential is contained in the Hamiltonian in Eq.~(\ref{eq:Hamiltonian}). As discussed in Ref. \citep{Thomson}, the wavefunction is not an eigenstate of the Hamiltonian and thus generates a complex energy expectation value (e.g. see Eq. (2) in Ref. \citep{Thomson}). Due to the large ratio between $r_0$ and $\sigma_0$, the exponential term is more than 20 orders of magnitude lower than the energy variance, and thus the complex term can be ignored. Moreover, the energy variance, $\Delta E_0$ = 0.1 keV, for the parameters in Table \ref{table:Variables_O_U}. This variance is several orders of magnitude smaller than the collision energies being investigated, therefore, providing good accuracy for the transmission coefficients.

\begin{figure}
    \centering
    \hspace*{-0.4cm}
    \includegraphics[scale=0.57]{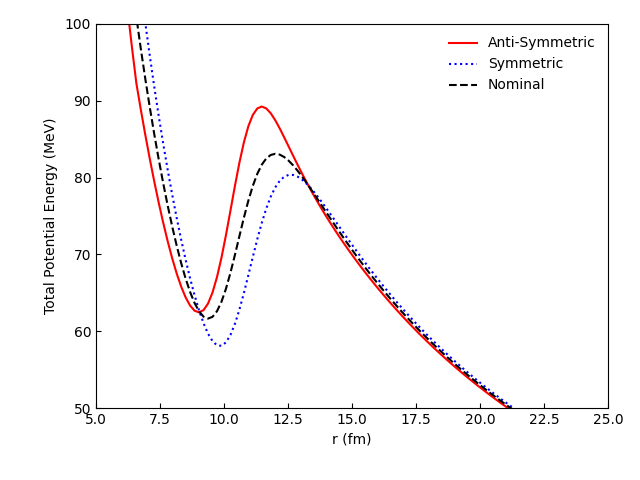}
    \caption{Comparison of the total bare interaction potential for $^{16}$O + $^{238}$U (dashed black line) and the eigenvalues of the total two-channels coupling potential for the anti-symmetric (solid red line) and symmetric (dotted blue line) eigenstates. The parameters of the total bare potential in Eq. (\ref{eq:potential}) are given in Table \ref{table:Variables_O_U}, the nominal Coulomb barrier being similar to that from the Sao Paulo double-folding potential model \citep{SaoPaolo}.}
    \label{fig:EigenChannels}
\end{figure}

The monopole potential, $V(r)$, in Eq.~(\ref{eq:Hamiltonian}) is taken as the sum of the repulsive Coulomb potential and the attractive Woods-Saxon nuclear potential,
\begin{equation}
    V(r) = \frac{Z_tZ_pe^2}{r} - \frac{V_0}{1 + \exp[\frac{r - R_0}{a_0}]},
    \label{eq:potential}
\end{equation}
for target and projectile charge numbers $Z_t$ and $Z_p$ respectively, nuclear potential depth $V_0$, nuclear radius parameter $R_0$, and nuclear diffuseness $a_0$. The Coulomb potential of two strongly overlapping ions is very uncertain and could be more realistically described either by the Coulomb potential between a charged particle and a uniformly charged sphere or by other phenomenological approximations \citep{Devries1975,Basu1989}. However, the details of the Coulomb potential at short radii may be irrelevant, as the Coulomb barrier is mainly determined for weakly overlapping ions and the internal part of the Coulomb barrier is embedded in a strong absorption region to describe the fusion. The imaginary potential, $iW(r)$, ensures an irreversible fusion process, where
\begin{equation}
    W(r) = -\frac{W_0}{1 + \exp[\frac{r - R_{i}}{a_{i}}]},
    \label{eq:ImaginaryPotential}
\end{equation}
and $W_0$, $R_i$ and $a_i$ are the corresponding potential depth, radius parameter and diffuseness for the absorption potential respectively. The additional sum in Eq. (\ref{eq:TDSE}) introduces the coupling potential matrix, $V_{nm}$. One can find the origin of these couplings in Ref. \citep{CCFULL}, the application of which is discussed in Ref. \citep{Vockerodt2019}. We therefore take the resultant potentials for the nuclear and Coulomb couplings for a rigid rotor in the $^{238}$U target. The nuclear coupling potential introduces a dynamical operator $\hat{O}$, into the Woods-Saxon potential \citep{CCFULL,Vockerodt2019},

\begin{equation}
    V_N(r, \hat{O}) = -\frac{V_0}{1 + \exp[(r - R_0 - \hat{O})/a]}.
    \label{eq:NuclearCouplingBase}
\end{equation}
The operator $\hat{O}= \beta_2R_T Y_{20}(\theta) + \beta_4R_T Y_{40}(\theta)$, where $\theta$ is the orientation angle of the symmetry axis of the deformed target nucleus relative to the internuclear axis. It can be represented by the matrix elements,
\begin{equation}
\begin{split}
    \hat{O}_{IJ} = & \sqrt{\frac{5(2I+1)(2J+1)}{4\pi}}\beta_2R_T\begin{pmatrix}
    I & 2 & J\\
    0 & 0 & 0
    \end{pmatrix}^2 + \\ & \sqrt{\frac{9(2I+1)(2J+1)}{4\pi}}\beta_4R_T\begin{pmatrix}
    I & 4 & J\\
    0 & 0 & 0
    \end{pmatrix}^2,
\end{split}
\label{eq:RotationalOperator}
\end{equation}
where $I$ and $J$ are the spins of the coupled rotational states of the target for quadrupole deformation parameter $\beta_2$, hexadecapole deformation parameter $\beta_4$, and $R_T = r_{coup}A_T^{1/3}$ where $r_{coup}$ is the coupling parameter. The eigenvalues and eigenvectors of the dynamical operator can then be calculated by,
\begin{equation}
    \hat{O}|\alpha\rangle = \lambda_\alpha |\alpha\rangle 
    \label{eq:EigenValues}
\end{equation}
and inserted into,
\begin{equation}
    V_{nm}^{(N)}(r) = \sum_{\alpha}\langle n|\alpha\rangle\langle\alpha|m\rangle V_N(r, \lambda_\alpha) - V_N^{(0)}(r)\delta_{n,m}
    \label{eq:NuclearMatrixElement}
\end{equation}
for the $|n\rangle = |I0\rangle$ and $|m\rangle  = |J0\rangle$ states of the ground rotational band of the $^{238}$U target \citep{CCFULL}. Eq. (\ref{eq:NuclearMatrixElement}) stems from the calculation of the matrix elements of the nuclear coupling potential in Eq. (\ref{eq:NuclearCouplingBase}) between the states of the ground rotational band of the target using a matrix algebra \citep{Kermode1993}. The resultant Coulomb couplings can be expressed directly as,
\begin{equation}
\begin{split}
    V_{nm}^{(C)}(r) = &\frac{3Z_pZ_t}{5}\frac{R_T^2}{r^3}\sqrt{\frac{5(2I+1)(2J+1)}{4\pi}}\\ &\times (\beta_2 + \frac{2}{7}\sqrt{\frac{5}{\pi}}\beta_2^2)\begin{pmatrix}
    I & 2 & J\\
    0 & 0 & 0
    \end{pmatrix}^2 + \\ &\frac{3Z_pZ_t}{9}\frac{R_T^4}{r^5}\sqrt{\frac{9(2I+1)(2J+1)}{4\pi}}\\ &\times (\beta_4 + \frac{9}{7\sqrt{\pi}}\beta_2^2)\begin{pmatrix}
    I & 4 & J\\
    0 & 0 & 0
    \end{pmatrix}^2, 
\end{split}
\label{eq:RotationalCoulomb}
\end{equation}
where the total coupling potential is the sum of the nuclear and Coulomb couplings, $V^{(N)}_{nm}$ and $V^{(C)}_{nm}$.  The resultant eigenvalues of the total coupling matrix can be seen in Fig. \ref{fig:EigenChannels}, and the associated eigenstates represent decoupled eigenchannels with a weighting probability factor of 50\%. The contribution of the symmetric eigenchannel with the lowest Coulomb barrier (dotted blue line) enhances the subbarrier fusion cross sections \citep{Balantekin}.

\begin{figure}
    \centering
    \hspace*{-0.4cm}
    \includegraphics[scale=0.57]{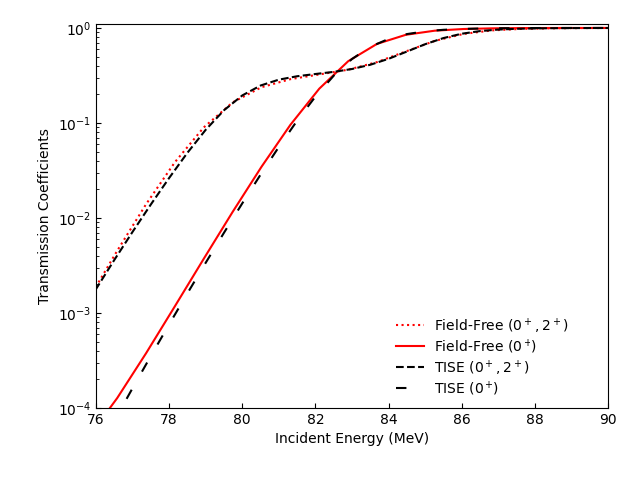}
    \caption{Comparison of the single-channel (solid red line) and coupled-channels (red dotted line) dynamic approach to fusion with the TISE approaches for a central ($l=0$) $^{16}$O + $^{238}$U collision.}
    \label{fig:SC_CC}
\end{figure}

We follow the time evolution of the wavepacket in small time-step $\Delta t$ applying the unitary transformation \citep{Tannor},
\begin{equation}
    \psi(r,t+\Delta t) = U(t + \Delta t, t) \, \psi(r,t),
    \label{eq:Evolution}
\end{equation}
where 
\begin{equation} 
    U(t + \Delta t, t) \approx \frac{\left(1 - \frac{i\hat{{\cal H}}\Delta t}{2\hbar}\right)}{\left(1 + \frac{i\hat{{\cal H}}\Delta t}{2\hbar}\right) },
    \label{eq:Unitary1}
\end{equation}
using the Crank-Nicolson formula. The state $\psi(r,t)$ represents the coherent linear superposition of the states $\psi_n(r,t)$ in Eq. (\ref{eq:TDSE}). The transmission coefficients for a central $^{16}$O + $^{238}$U fusion interaction with and without a single excited state can be seen in Fig. \ref{fig:SC_CC} for the parameters in Table \ref{table:Variables_O_U}, calculated using $\mathcal{T}_{l=0}(E_0) = 1 - \langle \psi (t_f) | \psi (t_f)  \rangle$ after a long propagation time $t_f$, as discussed in Ref. \citep{Thomson}. These calculations are compared to those calculated by solving the time-independent Schr\"odinger equation (TISE) with the incoming wave boundary condition and the nominal total interaction potential (dashed black line) in Fig. \ref{fig:EigenChannels}. The static coupled-channels calculations are carried out using the modified Numerov method, as implemented in Ref. \citep{CCFULL}. This comparison demonstrates the good accuracy of the quantum dynamical coupled-channels calculations.

\begin{table}
\centering
\caption{Model variables for the $^{16}$O + $^{238}$U nuclear fusion reaction. Both the nuclear and absorption radius parameters, $R_0$ and $R_i$, should be multiplied by the factor $A_O^{1/3} + A_U^{1/3}$. Deformation parameters are taken from Ref. \citep{238Udeformation}.}
\begin{tabular}{ p{2cm} p{2cm} p{4.3cm}}
\multicolumn{3}{c}{} \\
    \hline
    Variable & Value & Description\\
    \hline
    $A_O$&  16 & Projectile mass number \\
    $Z_O$&  8 & Projectile proton number\\
    $A_U$& 238 & Target mass number\\
    $Z_U$ & 92 & Target proton number\\
    $\Delta r$&  0.5 fm & Spatial step\\
    $r_0$   & 600 fm& Initial wavepacket position\\
    $r_{max}$ & 900 fm & Maximum separation \\
    $\sigma_0$ & 80 fm & Wavepacket spatial width\\
    $\Delta t$ & 0.01 zs & Time step (1 zs = $10^{-21}$ s)\\
    $R_0$ & 1.2035 fm & Nuclear radius parameter\\
    $a_0$ & 0.64 fm & Nuclear diffuseness  \\
    $V_0$ & 60.4 MeV & Woods-Saxon potential depth\\
    $R_i$ & 1.07 fm & Absorption radius parameter\\
    $a_i$ & 0.3 fm & Absorption diffuseness \\
    $W_0$ & 500 MeV& Absorption potential depth\\
    $r_{coup}$ & 1.06 fm & Coupling parameter\\
    $\epsilon_{2^+}$ & 44.92 keV & First excited state of $^{238}$U\\
    $\beta_2$ & 0.289 & Quadrupole deformation parameter for $^{238}$U\\
    $\beta_4$ & 0.124 & Hexadecapole deformation parameter for $^{238}$U\\      
 \hline
\end{tabular}
\label{table:Variables_O_U}
\vspace*{0.4cm}
\end{table}

\subsection{Laser-nucleus interaction}
We follow Refs. \citep{Wenjuan, Semi-Classical} and redefine the system Hamiltonian by introducing a laser-nucleus interaction potential, $V_{int}(r,t)$, on the diagonal of Eq. (\ref{eq:CouplingHamiltonian}) for a linearly polarised monochromatic laser field. 
$V_{int}(r,t)$ is given in the dipole approximation by
\begin{equation}
    V_{int}(r, t) = -\frac{e}{\mu} Z_{eff} \vec{A}(t) \, \hat{p} + \frac{Z^2_{eff}}{2 \mu}{\vec{A}}^2(t),
    \label{eq:Interaction}
\end{equation}
where $\hat{p}$ is the radial linear momentum operator, and $Z_{eff}$ is the effective charge,
\begin{equation}
    Z_{eff} = \frac{Z_OA_U - Z_UA_O}{A_O+A_U},
    \label{eq:Effective}
\end{equation}
and $\vec{A}(t)$ is the vector potential of the laser field,
\begin{equation}
    \vec{A}(t) = A_0 F(t)\cos(\omega t + \delta) \, \vec{e}_r,
    \label{eq:Field}
\end{equation}
for field strength $A_0 = \sqrt{2I/(\epsilon_0 \, c \, \omega^2)}$, intensity $I$, frequency $\omega$, phase factor $\delta$ and pulse shape function $F(t)$. In the calculations below, if not stated otherwise, we consider $\delta=0$, $\hbar \omega= 1$ eV, and $I = 10^{28}$ Wcm$^{-2}$. These parameters provide a large optimistic value of $A_0$ with currently achievable lasers \citep{Wenjuan}, maximising the effects of the laser-nucleus interaction on fusion. The strength of such effects declines when the angle between the laser's polarisation axis and the internuclear radius increases towards $\pi/2$ \citep{Wenjuan, Semi-Classical}. Introducing laser pulses into the system involves selecting a non-unit pulse shape function, $F(t)$, in Eq.  (\ref{eq:Field}). For the purpose of this interaction, we choose the common pulse shape,
\begin{equation}
    F(t) = \sin^2(\omega_p t)
\end{equation}
where $\omega_p$ is the pulse frequency. 

In a coupled-channels regime, the relevant off-diagonal laser-nucleus couplings of the Hamiltonian can be found in Ref. \citep{Palffy_2008} given by,
\begin{equation}
\begin{split}
    \langle I_e,M_e|\hat{\mathcal{H}}_I|I_g,M_g\rangle \approx & \hspace{0.1cm} E_k e^{-i\omega_kt}\sqrt{2\pi}\sqrt{\frac{L+1}{L}}\frac{k^{L-1}}{(2L+1)!!}\\
    & \times C(I_e I_g L;M_e - M_g - \sigma_p) \\
    & \times \sqrt{2I_g+1}\sqrt{B(\lambda L,I_g \rightarrow I_e)},
\end{split}
\label{eq:LaserCouplings}
\end{equation}
where $L$ is the photon multipolarity for a nuclear transition of type either electric ($\lambda=\mathcal{E}$) or magnetic ($\lambda=\mathcal{M}$), $I_g$ and $I_e$ are the angular momenta for their respective states, $M_g$ and $M_e$ are their respective magnetic sublevels, $E_k$ and $\omega_k$ are the electric field amplitude and laser frequency for a given wavenumber $k$ respectively, $\sigma_p$ is the polarisation, $C(j_1 \, j_2 \, j_3; m_1 \, m_2 \, m_3)$ denotes the relevant Glebsch-Goldan coefficient, and $B(\lambda L,I_g \rightarrow I_e)$ is the reduced transition probability. For the $^{238}$U target, the reduced electric quadrupole transition probability between the ground state and the first excited state is $B(\mathcal{E} 2,I_g \rightarrow I_e)=12.09 \times 10^4$ e$^2$fm$^4$ \citep{Palffy_2008}. In Eq. (\ref{eq:LaserCouplings}), the double factorial $n!! = n(n-2)(n-4)...$ down to 1 for an odd value of $n$ and 2 for an even value of $n$ \citep{Palffy}.

\subsection{Fusion Cross-Section}
\begin{figure}
    \centering
    \hspace*{-0.4cm}
   \includegraphics[scale=0.57]{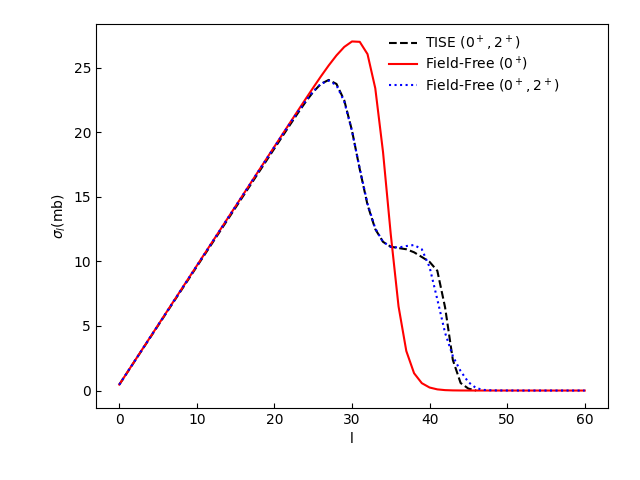}
    \caption{Fusion cross-sections for an $^{16}$O + $^{238}$U collision with varying initial angular momenta $l$ and $E_0=90$ MeV. We compare these cross-sections for both single (solid red line) and coupled-channel (blue dotted line) dynamic calculations with a TISE (dashed black line) approach.}
    \label{fig:Cross}
\end{figure}

\begin{figure}
    \centering
    \hspace*{-0.4cm}
    \includegraphics[scale=0.57]{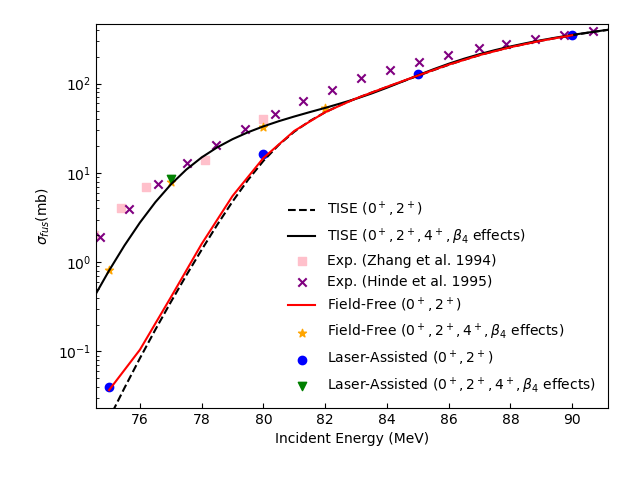}
    \caption{Total cross-section for an $^{16}$O + $^{238}$U coupled-channel fusion reaction with the addition of a laser field (blue dots) compared with the time-independent fusion cross-sections (dashed black line) and the fusion cross-sections without the presence of a laser field (solid red line). The total fusion cross-sections are shown for both a single excited state ($0^+, 2^+$) and 2 excited states ($0^+, 2^+, 4^+, \beta_4$ effects). The field-free experimental cross sections are taken from Refs. \citep{Zhang1994,Hinde1995}. The laser parameters are $I=10^{28}$ Wcm$^{-2}$, $\hbar \omega = 1$ eV, and $F(t)=1$. The Coulomb barrier height is 83 MeV.}
    \label{fig:TotalCross}
    \end{figure}

The partial wave cross-sections can be calculated for each orbital angular momentum $l$ \citep{Balantekin},

\begin{equation}
    \sigma_l(E_0) = \frac{\pi\hbar^2}{2\mu E_0}(2l+1)T_l(E_0).
\end{equation}
The total fusion cross-section for a given initial average energy $E_0$ can then be calculated by the sum of each partial wave cross-section,
\begin{equation}
    \sigma_{fus}(E_0) = \sum_l \sigma_l(E_0).
\end{equation}
The partial wave cross-sections for an $^{16}$O + $^{238}$U interaction with initial average energy $E_0 = $ 90 MeV can be seen in Fig. \ref{fig:Cross}. Both single-channel and coupled-channel simulations are compared using $l =$ 60 as a cut-off point, in which the total contributions of the subsequent partial waves are lower than 1/1000th of the total fusion cross-section. 
The total fusion cross-section in Fig. \ref{fig:TotalCross} is calculated for the coupled-channels regime both with (blue dots) and without (red solid line) the assistance of a laser field. The results for the 2 additional excited states are given by the orange stars and single green triangle. Given the significant increase in numerical calculations required for generating the fusion cross-sections, only a few specific energies have been selected for the laser-assisted fusion cross-sections.
The total fusion cross-section results mimic those found in Figs. \ref{fig:SC_CC} and \ref{fig:Laser} in which the field-free scenario aligns well with the time-independent approach and agrees fairly well with observations. For numerical simplicity, the octupole vibration in $^{238}$U was neglected. Again, the laser-assisted regime sits close to the field-free scenario.

\begin{figure}
    \centering
    \hspace*{-0.4cm}
    \includegraphics[scale=0.57]{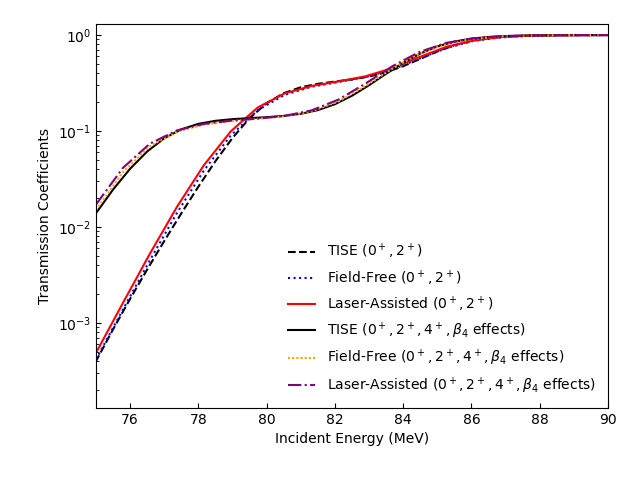}
    \caption{Transmission coefficients for $l=0$ in an $^{16}$O + $^{238}$U fusion reaction both with and without a laser field (red solid and blue dotted lines, respectively) in the coupled-channels regime, with varying initial average collision energies, compared with the TISE calculations (black dashed line).
    The transmission coefficients are shown for both a single excited state ($0^+, 2^+$) and 2 excited states ($0^+, 2^+, 4^+, \beta_4$ effects). The laser parameters are $I=10^{28}$ Wcm$^{-2}$, $\hbar \omega = 1$ eV, and $F(t)=1$. The Coulomb barrier height is 83 MeV.}
    \label{fig:Transmission}
\end{figure}

\section{Results and discussion}
\begin{figure}
    \centering
    \hspace*{-0.4cm}
    \includegraphics[scale=0.58]{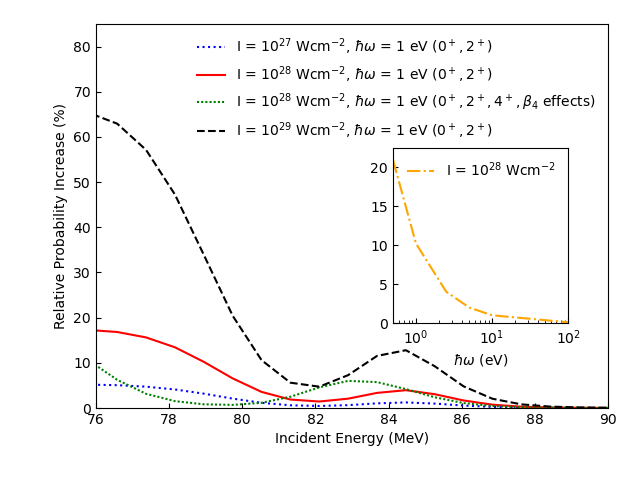}
    \caption{The percentage enhancement of the subbarrier $^{16}$O + $^{238}$U fusion probabilities for $l=0$ due to the laser-nucleus interaction for different values of intensity and photon's energy of a quasi-static ($F(t)=1$) laser field. The latter is shown in the inset for $E_0=79$ MeV. The dotted green line shows the fusion enhancement including the $4^+$ excitation with $\beta_4$ effects. The Coulomb barrier height is 83 MeV.}
    \label{fig:Laser}
\end{figure}

The transmission coefficients affected by the laser field are shown in Fig. \ref{fig:Transmission} for both a single $2^+$ excited state (red solid line) and 2 additional $2^+$ and $4^+$ excited states (purple dash-dot line).
Fig. \ref{fig:Transmission} shows that the laser field enhances the fusion probabilities. Fig. \ref{fig:Laser} focuses on the enhancement of the probabilities, showing the percentage increase in transmission probability when introducing the laser-nucleus interaction. Much like the results found in Ref. \citep{Thomson}, a significant enhancement is observed for increasing laser intensities. The unique properties of the coupled-channels interaction introduce an additional probability enhancement around $E_0=84$ MeV. From Fig. \ref{fig:EigenChannels} we can see that the symmetric and anti-symmetric eigenstates of the static coupling potential produce two effective Coulomb barriers (solid red and dotted blue lines). A radial oscillation in each of these barriers is caused by the quiver motion of the effective charge of the system \citep{Semi-Classical,Misicu_2013,Misicu_2019}, which is caused by the first term in Eq. (\ref{eq:Interaction}). The maximal amplitude of this radial \textit{Zitterbewegung}, $\alpha_0 = e Z_{eff} A_0/\mu \omega$, increases with increased intensity and decreased frequency of the laser field.  In the
Kramers-Henneberger representation of the wavepacket \citep{Henneberger}, the effect of the diagonal laser-nucleus interaction is fully transferred into the argument of both the absorptive potential and the eigenbarriers shown in Fig. 1, e.g., $V(r-\alpha_0 \sin \omega t)$. The eigenbarriers would be radially shaking, which allows the potential pocket to capture more probability from the incident wavepacket. However, the enhancement seen due to the off-diagonal laser-nucleus couplings in Eq. (\ref{eq:LaserCouplings}) is negligible with a maximum additional enhancement of the order of $10^{-9}$ $\%$. Those off-diagonal laser-nucleus couplings are dwarfed by the static coupling potentials provided by Eqs. (\ref{eq:NuclearMatrixElement}) and (\ref{eq:RotationalCoulomb}). Therefore, the meaningful fusion enhancements are as a result of the diagonal laser-nucleus interaction terms. This laser-assisted fusion enhancement is maximal when the laser's polarization axis is aligned with the internuclear axis.  

\begin{figure}
    \centering
    \includegraphics[scale=0.57]{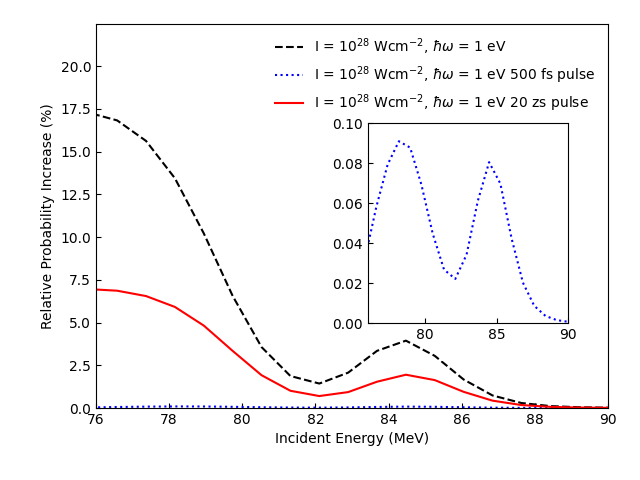}
    \caption{Percentage increase in transmission probability for $l=0$ due to the laser-nucleus interaction for an $^{16}$O + $^{238}$U coupled-channels reaction with different laser pulse duration. The inset focuses on the 500 fs ultrashort pulse.}
    \label{fig:RelativePulse}
\end{figure}

Fig. \ref{fig:RelativePulse} shows that for a typical ultrashort pulse of 500 fs (blue dotted line) we have not reached the necessary laser amplitude to generate a significant quiver amplitude. Since the interaction takes place over a timescale of the order of 100 zs, a pulse duration of 500 fs will have not changed over the course of the interaction and therefore have little effect on the system. A 20 zs pulse (red solid line) has been included to highlight the pulse duration required to reach the effects of the quasi-static laser field. Due to the large time period of the laser field relative to the interaction time, the continuous laser field can be considered quasi-static. 

Current laser technology is insufficient to produce significant fusion enhancement. For example, using the parameters of the Vulcan laser at the Central Laser Facility in the UK \citep{Vulcan} ($I=10^{23}$ Wcm$^{-2}$, $\hbar \omega = 1$ eV, and short pulses of $500$ fs) the largest fusion enhancement for $^{16}$O + $^{238}$U is 0.05\%. However, with the rapid advancements in laser power, reaching the fusion enhancements of the present work may eventually be achievable.

\section{Summary} 
We have presented a quantum dynamical approach to coupled-channels heavy-ion fusion for an $^{16}$O + $^{238}$U collision. It was successfully tested against the solution of the time-independent Schr\"odinger equation. The present fusion model describes the average trend of the $^{16}$O + $^{238}$U fusion excitation function. This average fusion baseline can be significantly enhanced by effects of a laser-nucleus interaction at subbarrier energies using very intense quasi-static lasers with low frequency. Furthermore, the channel couplings introduce a significant enhancement around the barrier energy, creating a secondary maxima. Laser pulses were shown to limit fusion enhancement. Femtosecond laser pulses significantly reduce the fusion enhancement, while zeptosecond laser pulses are required to reach the enhancements found using quasi-static laser fields.

\textbf{Acknowledgements.} This work was supported by the United Kingdom Science and Technology Facilities Council (STFC) under Grant No. ST/W5078421/1 and ST/Y000358/1, and by the Leverhulme Trust (UK) under Grant No. RPG-2019-325. 

  \bibliographystyle{ieeetr}
  \renewcommand\refname{\centering References \\ \hrulefill \\}
  \bibliography{references}

\end{document}